\documentclass[letter]{aa} 
\usepackage{graphicx}
\usepackage{txfonts}
\usepackage{subfig}
\begin{document}
\title{A long-period Cepheid variable in the starburst cluster 
VdBH222
\thanks{Based on observations made at the European Southern Observatory, 
Paranal, Chile under programs ESO 093.D-0168} 
}
\author{J.~S.~Clark\inst{1}
\and I.Negueruela\inst{2}
\and M.~E.~Lohr\inst{1}
\and R.~Dorda\inst{2}
\and C.Gonz\'{a}lez-Fern\'{a}ndez\inst{3}
\and F.~Lewis\inst{4,5}
\and P.~Roche\inst{5}}
\institute{
$^1$Department of Physics and Astronomy, The Open 
University, Walton Hall, Milton Keynes, MK7 6AA, UK\\
$^2$Departamento de F\'{i}sica, Ingenier\'{\i}a de Sistemas y Teor\'{i}a 
de la Se\~{n}al, Universidad de Alicante, Apdo. 99,
E03080 Alicante, Spain\\
$^3$Institute of Astronomy, University of Cambridge, Madingley Road, 
Cambridge CB3 0HA, UK\\
$^4$Astrophysics Research Institute, Liverpool John Moores University, 146 Brownlow Hill, Liverpool L3 5RF, UK\\
$^5$School of Physics \& Astronomy, Cardiff University, The Parade, CF24 
3AA, Cardiff, UK}

   \abstract{Galactic starburst clusters play a twin role in astrophysics, 
serving as  laboratories for the study of stellar physics and 
also delineating the structure and recent star formation history of the 
Milky Way.}
{In order to exploit these  opportunities we have undertaken a 
spectroscopic  survey of the  red supergiant dominated  young massive 
clusters thought to be present at both near and  far ends of the Galactic Bar.}
{Specifically,  multi-epoch observations were employed to 
identify and investigate 
stellar  variability and its potential role in initiating mass loss 
amongst the cool super-/hypergiant populations of these aggregates.}
{Significant spectroscopic variability suggestive of   
radial pulsations was found for the yellow supergiant VdBH 222 
\#505. Follow-up photometric investigations revealed  modulation 
with a period of $\sim23.325$~d; both timescale and pulsational profile
are  consistent with a Cepheid classification.}
{\#505 is one of the longest period Galactic cluster Cepheids identified 
to date and  hence of considerable use in 
constraining the bright end of the period/luminosity relation at solar metallicities. In conjunction 
with extant photometry we infer a distance of $\sim6$kpc for VdBH222 and an  age of 
$\sim20$Myr. This 
results in a moderate reduction in both the integrated cluster mass ($\sim2\times10^4M_{\odot})$ and the initial masses of  the evolved cluster members ($\sim10M_{\odot}$). As such VdBH222 becomes 
an excellent test-bed for studying the properties of some of the lowest mass stars observed 
to undergo type-II supernovae. Moreover, the distance is in tension with 
a location of VdBH 222 at the far end of the Galactic Bar. Instead a birthsite in the near 3kpc 
arm is suggested; providing compelling 
evidence of  extensive  recent star  formation in a region of  the inner 
Milky Way which has hitherto been thought to be devoid of such activity.}


\maketitle

\section{Introduction}

 Starburst or young massive clusters 
and  the wider cluster complexes or stellar  associations within 
which they are embedded are ubiquitous in external galaxies and 
appear to be a natural consequence of the 
physical processes that govern star formation. Historically such entities 
had been under-represented within our own Galaxy due to heavy extinction 
within the  plane. However the increased prevalence of 
large-scale (near-)infrared surveys has led to a realisation that 
it does indeed host a rich population of such stellar aggregates 
($>10^4M_{\odot}$, $<20$Myr; cf. Clark et al. \cite{clark13}).
The implications of such a population is two-fold: they provide a 
detailed record of recent (high-mass) star formation activity within the 
Galaxy 
and delineate its structure, while their  
constituent co-eval stellar populations make them ideal test-sites for 
the study of stellar evolution.  

Over the past decade it has become clear that within the last 
$\sim10-20$Myr 
a major episode of star 
formation  occurred at the intersection of the base of the Scutum-Crux 
arm with the near end of the Galactic Bar, resulting in the emergence of 
at least six young massive clusters embedded within extended halos/associations. These are 
delineated by their rich red supergiant (RSG) 
populations and  have  integrated masses of $\gtrsim10^4M_{\odot}$ 
(RSGC1...6; Figer et al. \cite{figer}, Davies et al. \cite{davies07}, 
Clark et al. 
\cite{clark09a}, Gonz\'{a}lez-Fern\'{a}ndez \& Negueruela \cite{gonzalez}, Negueruela et al. \cite{negueruela10}, 
\cite{negueruela11}, \cite{negueruela12}). An immediate question raised by this 
finding is 
whether similar activity is present at the far end of the Galactic Bar. 
Subsequent (ongoing) observations revealed 4 potential candidates - 
Mercer 81 (Davies et al. \cite{davies12}), [DBS2003] 179 (Borissova et al. 
\cite{borissova08}, \cite{borissova12}), VdBH 222 (Marco et al. 
\cite{marco}; Ma14) and Teutsch 85 (Marco et al. in prep.) - apparently located in this 
vicinity. Despite comparable masses, both Mercer 81 and [DBS2003] 179 
appear significantly younger ($\sim2-5$Myr) than the remaining clusters, with  RSGC1...6, 
VdBH222 and Teutsch 85 all estimated to be $\sim10-20$Myr old. 

The rich cool super-/hypergiant populations of the later clusters permits 
an investigation of this  critical phase of  post-main sequence 
evolution. The extreme mass-loss rates of red supergiants
 (RSGs) and yellow super-/hypergiants (YS-/HGs) shape the evolutionary 
pathways and endpoints of massive stars and both types of star have been 
directly identified as the progenitors of type II supernovae 
(e.g. Georgy \cite{georgy}). In an effort to characterise
the  variability exhibited by cool super-/hypergiants (e.g. Clark et al. 
\cite{clark10}, de Jager \cite{dejager}, Kiss et al. 
\cite{kiss})  - and hence investigate  a possible association between this behaviour and 
(pulsationally driven?) enhanced mass-loss - in 2014 we initiated a multi-epoch 
spectroscopic survey of members of  RSGC2, RSGC3 and VdBH222, including the YSGs 
VdBH222 \#371 and \#505 (henceforth \#371 and \#505).

\begin{figure}
\includegraphics[width=6.5cm,angle=270]{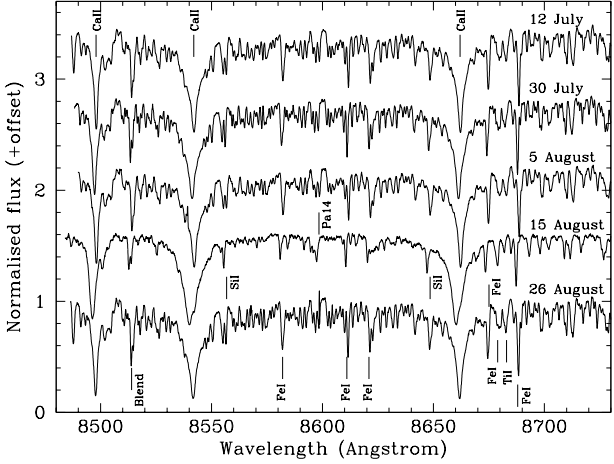}
\caption{Five epochs of 
$I$-band spectroscopy of \#505.
 The dates of individual observations are associated with the 
relevant spectra. Prominent transitions are indicated; note that Paschen 13, 15 \& 16 are located in
the blue wings of the Ca\,{\sc ii} absorption lines while the 
blend at 8514{\AA}  is dominated by   Fe\,{\sc i}. Each spectrum has been wavelength 
corrected to take into account the mean cluster velocity of VdBH222 of $\sim -99$kms$^{-1}$.
We highlight the relative weakness of the rich, low-excitation, metallic photospheric lines and greater 
widths of the Ca\,{\sc ii} transitions  in the spectrum 
obtained on August 15 in comparison to the remaining spectra.}
\end{figure}

\begin{figure}
\includegraphics[width=5.5cm,angle=0]{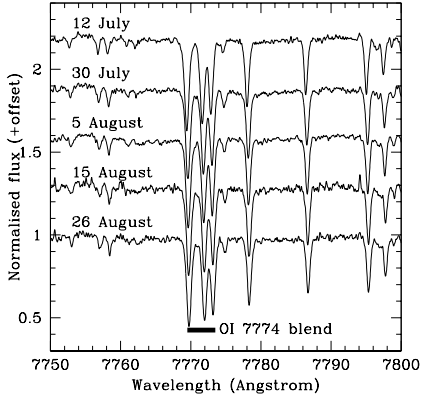}
\caption{Subsection of the spectra of \#371 encompassing the O\,{\sc i} 7774{\AA} blend 
as well as low excitation metallic photospheric lines, demonstrating a lack of RV variability which
appears to exclude identification as a Cepheid.}
\end{figure}

\section{Data acquisition, reduction and analysis}

\subsection{Spectroscopy}
During the 44~d interval between  2014 July 12  to August 26 we obtained 
five epochs of 
spectroscopy of   
\#505 (Fig. 1). The observations were made using the Fibre Large Array Multi Element 
Spectrograph
(FLAMES; Pasquani et al. \cite{pasquani}), located on VLT UT2 \textit{Kueyen} at Cerro
Paranal. The MEDUSA mode of the GIRAFFE spectrograph was employed and the 
setup HR21 covering the 8484-9001${\AA}$ range with $R = \lambda/\Delta\lambda\sim 16200$ was chosen, with 
individual integrations of 555s. The raw data were reduced with the GIRAFFE pipeline and have been de-biased, flat-fielded and wavelength-calibrated (including a correction to the helio-centric reference system). While discussed in detail  in a future publication, we simultaneously observed  \#371 with the UVES feed of FLAMES. Observations were made from $\sim6700-10130${\AA} and a resolving power of $\sim$47,000.

It is apparent that significant variability is present in both the spectral morphology and radial 
velocity (RV) of \#505. This is  reminiscent of the behaviour of the the cool hypergiant 
population within Westerlund 1 (e.g. Clark et al. \cite{clark10}) and hence we interpret 
this behaviour in terms of radial pulsations.  While all five spectra are dominated by 
photospheric lines from low excitation metallic species, these appear noticeably weaker in the  
the spectrum from August 15, suggesting an earlier spectral type at this epoch. 
Classification of stars in this spectroscopic 
window is described in detail in Negueruela et al. (\cite{negueruela11}, \cite{negueruela12}), 
Ma14 and refs. therein. Specifically, following Carquillat et al. 
(\cite{car}) we employ the Ti\,{\sc i} 8683{\AA} to  Fe\,{\sc i} 8679{\AA} 
line ratio  as the primary criterion to infer spectral types for our data. We suggest that \#505 
varies between  $\sim$G2 Ia ($\sim5000$K; August 15 and consistent with the original 
classification of Ma14) to $\sim$K0 Ia ($\sim4200$K; July 12 
and August 26). Note  the rapidity of the spectral evolution, with the transition between these 
extrema occurring over only 10~d. Utilising the strong Fe\,{\sc i} 
lines in the spectrum we determine a 
 maximum ${\Delta}RV{\sim}40$kms$^{-1}$ between the same two epochs of 
observations. The mean of the five measurements - 
$v\sim -95\pm3$kms$^{-1}$ - is within 1$\sigma$ of the mean cluster velocity 
($v\sim -99\pm4$kms$^{-1}$; Ma14) strongly supporting membership.

Given the correspondence between the temperatures inferred for \#505 and the instability strips 
defined by Tammann et al. (\cite{tammann}) and Bono et al.  (\cite{bono}), a Cepheid classification 
is immediately suggested, with the magnitude of ${\Delta}RV$ observed consistent with this 
hypothesis\footnote{Notwithstanding that the bolometric  luminosity estimated by 
Ma14 ($(L_{\rm bol}/L_{\odot})\sim4.6$ for a kinematic distance of 
$\sim10$kpc) exceeds the upper limit of the  Cepheid instability strip  reported by Anderson et al. 
(\cite{anderson14}).}. Following from the analysis of VdBH22 and constituent stars by Ma14, 
 application of the period/luminosity ($P/L$) or period/age relations 
for 
Cepheids would then suggest a periodicity for \#505 in the range 
$\sim40-50$~d (Bono et al. 
\cite{bono}, 
Storm et al. \cite{storm}).

Conversely, no evidence of pulsations, either in  terms of RV variability or changes in spectral 
morphology,   was identified between  the five UVES spectra of the second - and more luminous - 
 cluster YSG, \#371 (Fig. 2).

\subsection{Photometry}
Motivated by our spectroscopy, follow-up optical photometry was acquired
 with  the suite of 1-metre instruments that form part of the Las Cumbres Observatory
 Global Telescope (LCOGT) network; details of both network
 instrumentation and operations can be found in Brown et al. (\cite{brown}). 
Observations were made in a $\sim110$~d interval from  2015 February 3  in  $V-$ (58 epochs with 40s 
exposures), $R-$  (52 epochs with 20s exposures) and SDSS $i-$  (24 epochs with 20s exposures) broadband filters.

Initial bias subtraction and flat-fielding were performed via the automatic
pipeline process described by Brown et  al. (\cite{brown}). The images were then 
realigned and trimmed to a common  coordinate system and area and point spread 
function (PSF)-fitting photometry was  carried out using the IRAF package DAOPHOT.  
Owing to the varying quality of the  images (FWHMs ranging from $\sim2$ to 
$\sim6$ pixels), the fitting was performed  individually for each frame, using 3-6 
stars  located in or near the cluster and of comparable brightness to the targets of 
interest  to construct each PSF.  The coordinates of detectable cluster stars were 
found from  a single good image, and then used as the starting points for all other 
frames, to ensure that the same stars were measured in each case.

Light curves were obtained for both YSGs 
 as well as the  nine RSGs and eight blue cluster members
identified by Ma14. Unlike \#505,   \#371 and the cluster RSGs exhibited 
near-identical 
night-to-night trends indicative of a lack of variability, 
while the fainter blue stars were more variable and their magnitudes more uncertain.   
Therefore differential light curves were  constructed for \#505 using \#371 
as a comparison star and, 
independently, an  average of the RSGs as a composite check star (Fig. 3); both methods 
yielding results fully consistent with one another.
 The resultant lightcurves  reveal \#505 to be periodically  variable.
 Phase dispersion minimisation was therefore employed to determine a period of 
$\sim23.325\pm0.050$~d for the full dataset. The $V-$ and $R-$ band
data  were then calibrated with extant photometry from Ma14, yielding the 
final folded light curves presented in Fig. 3. Maxima were obtained 
on JD 2457138.449 ($V-$band) and 2457138.748 ($R-$band).

The dependence of pulsation profile on period in Cepheids is well known 
(the Hertzsprung progression; Hertzsprung \cite{hertz}, Soszy\'{n}ski 
et al. \cite{sosz}). Comparison of the pulsational profile of \#505 to 
that of T Mon (period $\sim27$~d; Freedman \& Madore \cite{freedman})
shows an encouraging correspondence; again entirely consistent with  a 
Cepheid classification.

\begin{figure}
\includegraphics[width=6cm,angle=0]{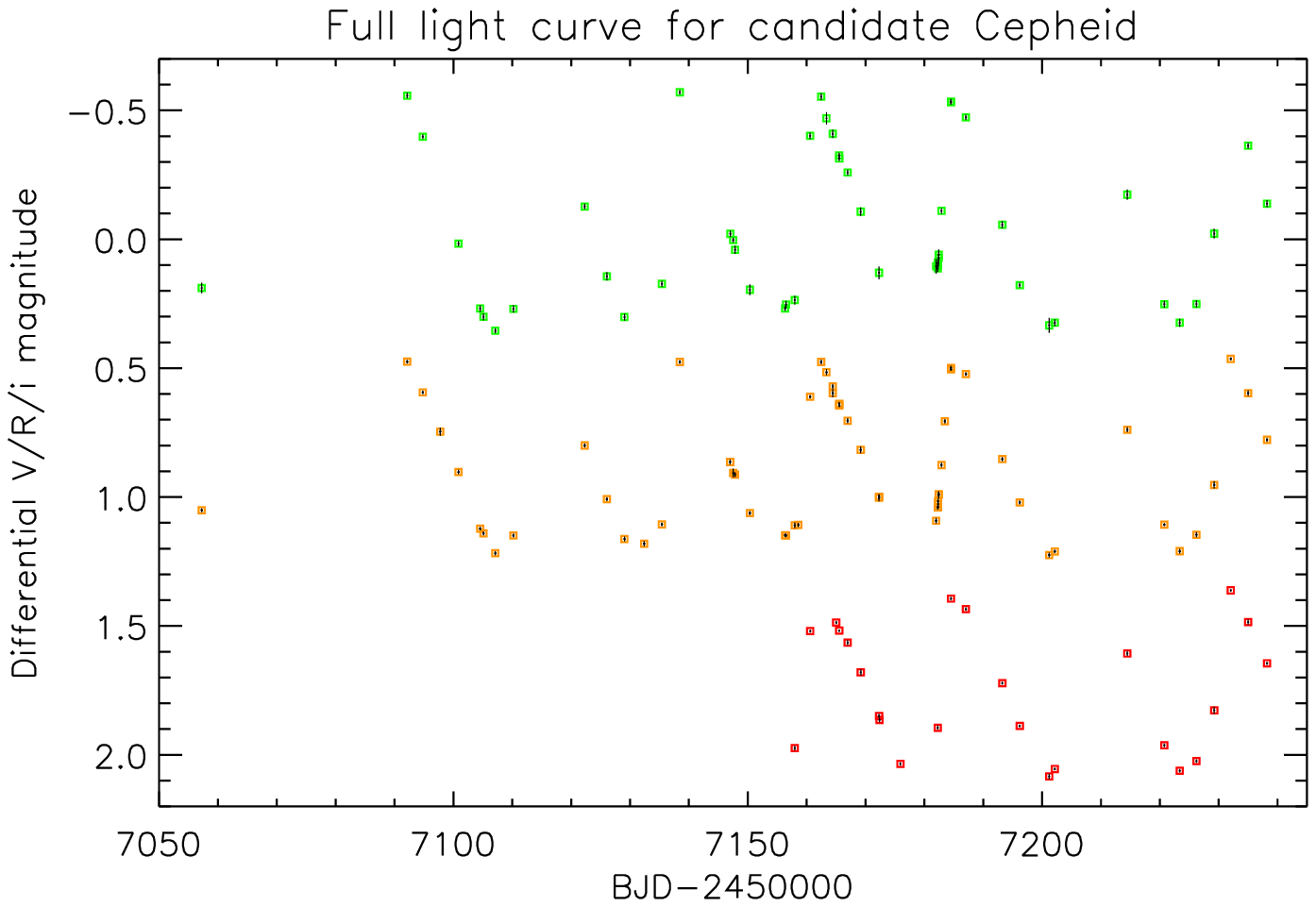}
\includegraphics[width=6cm,angle=0]{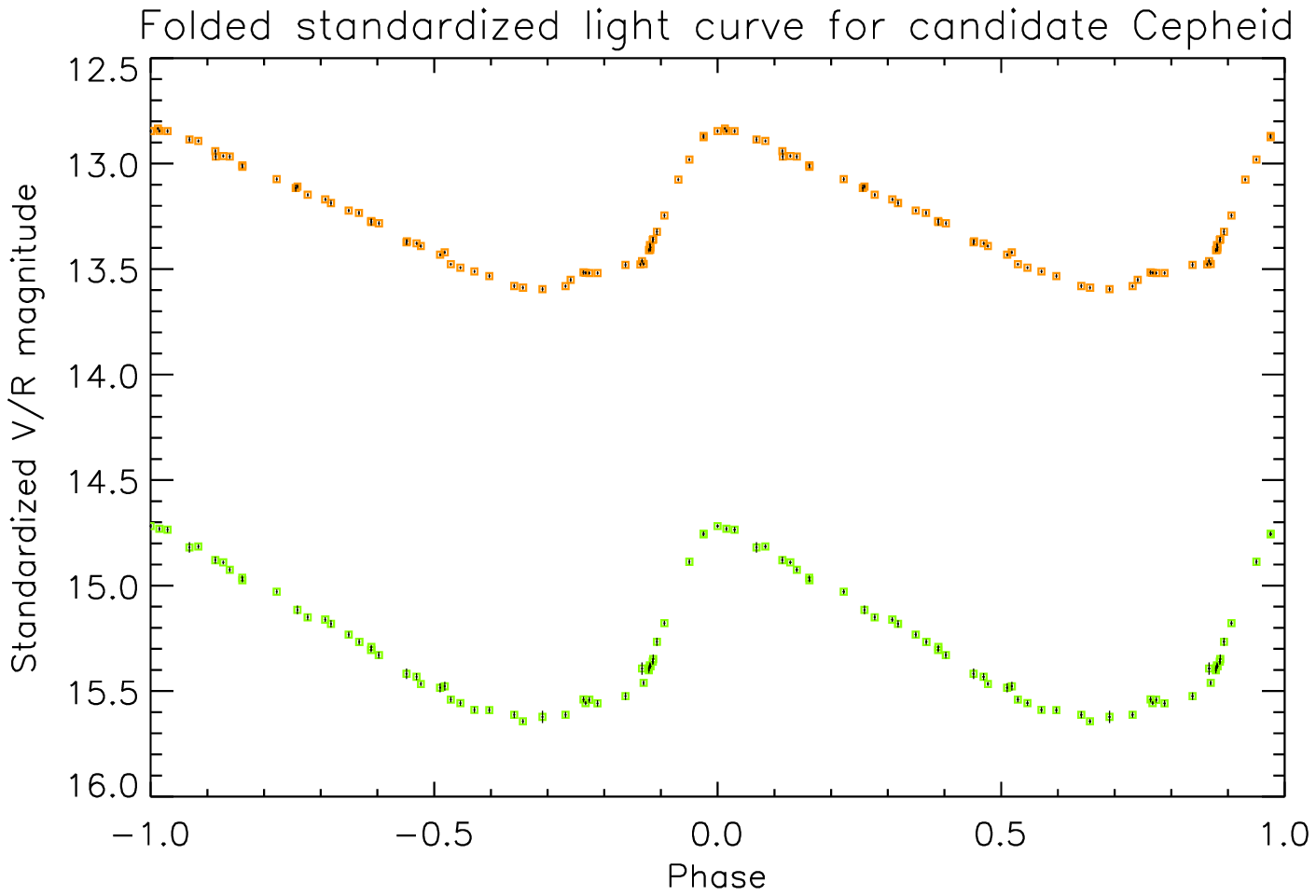}
\caption{{\bf Top panel:} Differential $V$-, $R$- and SDSS $i$- band lightcurves of VdBH \#505 (green, orange and red symbols respectively). Errors associated with the data  are smaller than the symbols.
{\bf
Bottom panel:}Calibrated $V$- and $R$-band lightcurves folded on our best-fit $\sim23.325\pm0.050$~d
pulsation period,  with phase zero set from the $V-$band maximum. Symbols and errors as above.}
\end{figure}

\section{Interpretation and concluding remarks}

 The location of \#505 at the core of VdBH222, as well as its placement on theoretical isochrones 
that accurately fit the location of the remaining RSG cohort (Ma14),  provide a compelling argument 
for cluster membership. As such the  importance of \#505 is two-fold. Firstly, it provides a direct 
measure of the distance to VdBH222.  Secondly, if an independent 
measure of the cluster distance is available - such as a kinematic determination from the mean 
cluster RV and/or isochrone  fitting to the evolved stellar population or the 
main sequence turn-off -  one may invert the argument and  utilise the observed  properties of \#505 to calibrate e.g. the 
P/L relationship for Cepheids.  Consequently, \#505 is particularly interesting  
since its period exceeds that of any of the cluster  Cepheids from the recent `optimal' data set 
compiled by  Anderson et al. (\cite{anderson13}) and is the second longest of the more recent census of Chen et al. (\cite{chen}); potentially providing a strong anchor point to the bright end of the P/L relation.

Utilising Eqn. 14 and 19 of Anderson et al. (\cite{anderson13}) we first determined 
$\langle{m_V}\rangle=15.22\pm0.07$ and, from  the pulsational period of  \#505, an absolute $V$-band 
magnitude $M_{V}=-5.0\pm0.3$ (log($L/L_{\odot})\sim4.1^{+0.2}_{-0.2}$;  bolometric correction from  
Levesque et al. \cite{levesque}).   The colour excess can be calculated from our data and the 
$J$-band magnitude from Ma14 via Eqns. 15 and 16 of Anderson et al. (\cite{anderson13}); we found 
$E(B-V)=2.07\pm0.07$. Adopting $R_V\sim2.9$ (Ma14) and the colours of a late G supergiant  (Fernie 
\cite{fernie}) yielded $A_v\sim6.5$ and hence a distance to \#505 of $5.8_{-0.7}^{+0.8}$kpc, noting 
that the formal errors quoted do not  take into account uncertainty in the determination of $R_V$ for 
the cluster (cf. Ma14).  Finally, employing the period-age relationship of Bono et al. (\cite{bono}) 
we arrived at  an age of $\sim24.5_{-4.1}^{+5.0}$Myr for the star.

Motivated by our current understanding  of  
Galactic structure and star formation activity,  Ma14 favoured an age of $\sim12$Myr
and distance  of 
$\sim10$kpc for VdBH222, placing it at the far end of the Galactic Bar. Such a location is also compatible with the far kinematic distance  inferred from the mean cluster RV. Nevertheless, such a solution is clearly in tension with the properties inferred from \#505; for example at such an age one would expect a pulsational period of $\sim69$~d (Bono et al. \cite{bono}). However the near kinematic distance of 
$\sim6$kpc (Ma14) is entirely  consistent with that determined {\em independently} for VdBH222 from \#505, with such a distance also consonant with  extant photometry, assuming an age of $\sim20$Myr (Ma14).

Although not a Cepheid, the spectrum of  \#371 provides one further  argument for the nearside distance to VdBH222. Our new spectroscopy encompasses the O\,{\sc i} 7774{\AA} blend (Fig. 2), from the strength of which 
(Equivalent Width$=-1.26\pm0.05${\AA}) we may infer  $M_V\sim-6.1\pm0.9$ 
following the calibration of Arellano Ferro et al. (\cite{arellano}). Assuming cluster membership this, too, is consistent with a distance of $\sim6$kpc. 

Before accepting this distance we recognise two potential {\em caveats}. Firstly, Ma14 identify  a population of blue cluster members of apparent B-spectral type, which they tentatively conclude to be luminosity class Ib-II. If VdBH222 is located at a distance of $\sim6$kpc these would instead be either main sequence or giant  stars; hence their properties provide an independent observational test of this revision. Secondly one might ask why the second cluster YSG, \#371, shows no spectroscopic or photometric evidence for pulsations - given that our new data are consistent  with with the  G0 Ia$^+$ classification of Ma14, implying log$T_{\rm eff}\sim3.72$ and hence potentially locating it   within the Cepheid instability strip (cf. Fig. 1 of Anderson et al. \cite{anderson14})? 
A possible explanation may be found in the proposed upper bound  for  the  instability strip  (log($L/L_{\rm bol})\sim4.3$; Anderson et al. \cite{anderson14}); even allowing for the nearside distance, the photometrically-determined luminosity of 
\#371 (Ma14) exceeds this value, while our spectroscopic estimate from the calibration of the O\,{\sc i} 7774{\AA} blend coincides with this limit (log($L/L_{\odot})\sim4.3^{+0.4}_{-0.3}$; consistent with the photometrically-determined value within the error estimates). In contrast, for a distance of $\sim6$kpc, the luminosity of   \#505 is below this threshold and hence one would expect it  to show Cepheid behaviour. Conversely, at a distance of 
 $\sim10$kpc both stars would exceed the upper luminosity limit of the Cepheid instability strip  and hence neither should exhibit pulsations (cf. footnote 1).  

 In such a scenario the discrepancy in luminosities between  \#371 and \#505 could result from the former being more massive and/or more rapidly rotating than the latter. Rapid rotation leads to a more luminous post main sequence 
evolutionary path for a given initial mass and also a longer lifetime on the main sequence (Anderson et al. \cite{anderson14}). The latter implies that under the assumption of 
co-evality for VdBH22, the progenitor of \#371 would have had to be more massive than \#505 for both stars to currently be observed in the same evolutionary phase, also implying  a greater luminosity for the 
former.

\subsection{Implications of a reduced distance to VdBH222}
What, then,  are the physical implications if VdBH222 is located at the nearside kinematic 
distance of $\sim6$kpc? In conjunction with extant photometry and the distribution of spectral types of known cluster members (Ma14) and given the relatively large  uncertainties associated with inferring an age from the pulsational period of \#505,  we favour a cluster  age of 
$\sim20$Myr. 
If confirmed via future observations of e.g. the main sequence turn-off, 
VdBH222 would be the oldest of the RSG dominated clusters (cf. Sect. 1) and its integrated  mass would
undergo a moderate downwards revision to $\sim2\times10^4M_{\odot}$ (Ma14, Clark et al. \cite{clark09b}).
Likewise the initial mass of the cluster members would be lower. 
Anderson et al. (\cite{anderson14}) quote a
 rotationally-dependent maximum mass of $\sim10.5-11.5M_{\odot}$ for Cepheids, which serves 
as a lower limit to the mass of \#371. Adopting the mass/luminosity relations of these 
authors suggests a slightly lower mass for \#505 of $\sim9.0^{+1.1}_{-1.1}M_{\odot}$, subject to uncertainties in distance, bolometric correction, reddening 
law and initial rotational rate. Estimates for both stars are therefore broadly consistent with that derived for the evolved stellar cohort from
the age of VdBH222.

This is of considerable interest since the evolutionary 
 endpoint of $\sim7-10M_{\odot}$ stars - e.g white dwarf, electron-capture or core-collapse supernova - is  uncertain (Poelarends et al. \cite{poel}, Jones et al. \cite{jones}, Doherty et al. \cite{doherty}). Observationally, searches for the pre-explosion counterparts of type-II supernovae return a progenitor mass-range of $\sim7-16M_{\odot}$ (Smartt et al. \cite{smartt}), suggesting that at least some such stars do explode. Unfortunately, to date few examples of low mass ($\lesssim10M_{\odot}$) RSGs within Galactic clusters have been identified; VdBH222 therefore provides an excellent laboratory for studying the properties of such objects. 
Likewise, \#505 and \#371 serve as ideal test-subjects to validate theoretical predictions of Cepheid behaviour for the type of high-luminosity star that is critical to extragalactic distance determinations 
(Anderson et al. \cite{anderson14}) and, ultimately, determination of the Hubble constant.

Finally, a distance of $\sim6$kpc is unexpected given our current understanding of Galactic structure and recent star formation history. Such a distance places VdBH222 in, or in the vicinity of, the inner 3kpc Galactic arm (van Woerden et al. \cite{vanW}). This is very poorly traced in 
molecular gas and historically has  not been thought to
host active star formation (e.g. Lockman \cite{lockman}). However methanol 
maser emission - considered to be uniquely associated with massive star 
formation - has recently been detected within it (Green et al. \cite{green}),
as have  both H\,{\sc ii} regions (Caswell \& Haynes \cite{caswell}) and 
IR-colour selected candidate massive young stellar objects (Busfield et al. 
\cite{busfield}), potentially challenging this  orthodoxy.

Sanna et al. (\cite{sanna}) placed the massive star-forming  region W31 in the 3kpc arm, although the findings of Furness et al. (\cite{furness}) dispute this assertion, favouring a much smaller distance. Given this uncertainty, 
 VdBH 222 provides  the first compelling observational proof of significant recent star-formation activity within this region. Indeed, by analogy to the RSG dominated clusters at the near-end of the Galactic Bar we might expect  VdBH222 to be associated with an equally  massive, diffuse association/halo, suggesting several $\sim10^4M_{\odot}$ of stars to have formed within the last $\sim20$Myr (a scenario we investigate in a future work). Nevertheless, and irrespective of this hypothesis, VdBH222 appears to represent the end-point of vigorous, recent  star formation in the 3kpc arm;  an entirely unexpected result given the apparent lack of sufficient raw material to fuel such activity and the dynamical effects of the Galactic Bar sweeping through the region.

\begin{acknowledgements}
We thank Hugo M. Tabernero for additional data reduction. 
 This research is partially
supported by the Spanish Ministerio de Econom\'{\i}a y Competitividad under
grant AYA2012-39364-C02-02, and the European Union.
The Faulkes Telescopes are maintained and operated by Las
Cumbres Observatory Global Telescope Network.

\end{acknowledgements}

{}
\end{document}